\documentclass[paper]{ieice}

\usepackage{cite}
\usepackage{insertfig}
\usepackage{times}
\usepackage{amsmath}
\usepackage{url}

\allowdisplaybreaks[1]

\makeatletter
\renewcommand{\thesection}{\@arabic\c@section}
\renewcommand{\thesubsection}{\thesection.\,\@arabic\c@subsection}
\makeatother

\renewcommand{\insertfigsize}{0.4}

\newcommand{\ds}{\displaystyle}

\usepackage{bm}
\newcommand{\vbf}[1]{\bm{#1}}
\newcommand{\hvbf}[1]{\bm{\hat{#1}}}
\newcommand{\tmp}[1]{{}^t#1}

\newcommand{\diff}{{\rm d}}
\newcommand{\diag}[1]{{\rm diag}(#1)}
\newcommand{\ex}[1]{{\rm E}(#1)}
\newcommand{\prob}[1]{{\rm Prob}\left[#1\right]}

\begin{document}

\title{Graph Degree Heterogeneity Facilitates Random Walker Meetings}
\authorlist{\authorentry[sakumoto@kwansei.ac.jp]{Yusuke Sakumoto}{m}{kwansei}
    \authorentry[ohsaki@kwansei.ac.jp]{Hiroyuki Ohsaki}{m}{kwansei}}

\titlenote{The preprint of this paper is available at \url{https://arxiv.org/abs/2005.11161}}
  
\affiliate[kwansei]{Kwansei Gakuin University, 2-1 Gakuen, Sanda, Hyogo 669-1337, Japan}

\maketitle

\begin{summary}
Various graph algorithms have been developed with multiple random
walks, the movement of several independent random walkers on a graph.
Designing an efficient graph algorithm based on multiple random walks
requires investigating multiple random walks theoretically to attain a
deep understanding of their characteristics.  The first meeting time
is one of the important metrics for multiple random walks.  The first
meeting time on a graph is defined by the time it takes for multiple
random walkers to meet at the same node in a graph.  This time is
closely related to the rendezvous problem, a fundamental problem in
computer science.  The first meeting time of multiple random walks has
been analyzed previously, but many of these analyses have focused on
regular graphs.  In this paper, we analyze the first meeting time of
multiple random walks in arbitrary graphs and clarify the effects of
graph structures on expected values.  First, we derive the spectral
formula of the expected first meeting time on the basis of spectral
graph theory.  Then, we examine the principal component of the
expected first meeting time using the derived spectral formula.  The
clarified principal component reveals that (a)~the expected first
meeting time is almost dominated by $n/(1+d_{\rm std}^2/d_{\rm
avg}^2)$ and (b)~the expected first meeting time is independent of the
starting nodes of random walkers, where $n$ is the number of nodes of
the graph.  $d_{\rm avg}$ and $d_{\rm std}$ are the average and the
standard deviation of weighted node degrees, respectively.  The
characteristic~(a) is useful for understanding the effect of the
graph structure on the first meeting time.  According to the revealed
effect of graph structures, the variance of the coefficient $d_{\rm
std}/d_{\rm avg}$~(degree heterogeneity) for weighted degrees
facilitates the meeting of random walkers.
\end{summary}

\begin{keywords}
First Meeting Time, Random Walk, Spectral Graph Theory
\end{keywords}

\section{Introduction}
\label{sec:intro}

Various graph algorithms have been developed with multiple random
walks, the movement of several independent random walkers on a graph,
as a result of graph algorithms' ease of analysis and light-weight
processing.  Notable applications include (a)~a search algorithm for
finding a particular node on a graph~\cite{Lv02:MRW,Gkantsidis:04:RW},
(b)~an algorithm for spreading information across graphs by exchanging
information only between adjacent nodes~\cite{Dutta15:Coalescing}, and
(c)~the rendezvous algorithm for efficient meeting of multiple random
walkers at the same node~\cite{Metivier00:Rendezvous}. Designing an
efficient graph algorithm based on multiple random walks requires
studying multiple random walks theoretically in order to understand
their characteristics at a deep level.

Several important metrics~(e.g., first hitting time, recurrence time,
cover time, re-encountering time, and first meeting time) have been
used for multiple random walks.  The first hitting time is the time it
takes for any random walker to arrive at a specified node, and it is
important for evaluating the performance of relevant search algorithms.
The recurrence time is the time required to return any one of the
random walkers to the starting node, and it is thus a particular case
of the first hitting time.  The cover time is the time it takes for any
random walker to reach all of the nodes and corresponds to the maximum
value of the first hitting times.  The cover time strongly affects the
information dissemination speed in the graph.  The re-encountering time
and the first meeting time are the times it takes for multiple random
walkers to meet at the same node.  The re-encountering time relates to
random walkers starting from the same node, and the first meeting time
relates to those starting from different nodes.  In particular, the
first meeting time is closely related to the rendezvous problem, a
fundamental problem in computer science.  The rendezvous problem occurs
in a number of engineering problems~(e.g., the self-stabilizing token
management system
problem~\cite{Israeli90:SelfStabilizing,Tetali91:SelfStabilizingRW}
and the $k$-server problem~\cite{Coppersmith93:OnLineAlgo}).  Designing
efficient algorithms for the rendezvous problem requires clarification
of the characteristics of the first meeting time.

The first meeting time of multiple random walks was analyzed
in~\cite{Aldous91:Meeting,Bshouty99:Meeting,Cooper09:MultiRW,Zhang14:Meeting,George16:Meeting}.
However, many of these previous studies focus on regular graphs.
In~\cite{George16:Meeting}, George et al. did pioneering work on
multiple random walks on non-regular graphs and derived a closed-form
formula for calculating the expected value of the first meeting time
in arbitrary graphs.  However, the effects of graph structures on the
expected first meeting time remain unclear.  Designing effective
algorithms using multiple random walks for realistic graphs~(e.g.,
social networks and communication networks) benefits from
understanding the effects of graph structures on the expected first
meeting time.  Since it is difficult to clarify these effects
numerically using the closed-form formula derived
in~\cite{George16:Meeting}, the effects must be examined using analysis
of multiple random walks.

In this paper, we analyze the first meeting time of multiple random
walks in arbitrary graphs and clarify the effects of graph structures
on its expected value.  First, we derive the spectral formula of the
expected first meeting time on the basis of spectral graph theory that
is used to analyze the characteristics of graphs.  Then, we examine
the principal component of the expected first meeting time using the
derived spectral formula.  The clarified principal component reveals
that (a)~the expected first meeting time is almost dominated by
$n/(1+d_{\rm std}^2/d_{\rm avg}^2)$, where $n$ is the number of nodes
in the graph and $d_{\rm avg}$ and $d_{\rm std}$ are the average and
standard deviation of weighted node degrees, respectively, and (b)~the
expected first meeting time is independent of the starting nodes of
random walkers.  Characteristic~(a) provides understanding of the
effect of the graph structure on the first meeting time.  In
addition, we verify the validity of the analysis results through
numerical examples.

The contributions in this paper are summarized as follows.
\begin{itemize}
\item We extend the analysis of a single random walks to multiple
  random walks using spectral graph theory.
\item We derive the spectral formula of the expected first meeting
  time.
\item We clarify the principal component of the expected first meeting
  time.
\item We reveal the effect of graph structures on the expected first
  meeting time.
\item We confirm the validity of the derived spectral formula and the
  clarified principal component for various networks with different
  scales and different structures.
\end{itemize}

The remainder of this paper is organized as follows. In
Sect.~\ref{sec:preliminary}, we describe the definition of graphs and
random walks and introduce the previous analysis of a single random
walk using spectral graph theory.  In Sect.~\ref{sec:analysis}, we
derive the spectral formula of the expected first meeting time using
spectral graph theory and clarify the principal component of the
expected first meeting time on the basis of the derived spectral
formula.  Sect.~\ref{sec:example} confirms the validity of the
analysis results through numerical examples.
Sect.~\ref{sec:conclusion} concludes the paper and discusses future
work.

\section{Preliminary}
\label{sec:preliminary}

In this section, we provide the definition of graphs and random walks
that we use in our analysis.  In addition, we review existing analysis
results of a single random walk based on spectral graph theory.

A graph is given by $G = (V, E)$, where $V$ and $E$ are a set of nodes
and a set of links, respectively.  Self-loop links $(i, i)$ for $i \in
V$ are not included in $E$.  The weight for link $(i, j) \in E$ is
$w_{ij}$, where $w_{ij} > 0$ and $w_{ij} = w_{ji}$.  We denote the set
of adjacent nodes of node $i \in V$ by $\partial i$. Letting $d_i$ be
the weighted degree of node $i \in V$, we define $d_i$ as
\begin{align}
 d_i := \sum_{k \in \partial i} w_{ik}.
\end{align}

We describe a random walk starting from node $a \in V$.  In this random
walk, the random walker at node $i \in V$ moves to adjacent node $j
\in \partial i$ using transition probability $p_{i \rightarrow j}$
given by
\begin{align}
  p_{i \rightarrow j} = \frac{w_{ij}}{d_i}.
  \label{eq:def_trans_prop}
\end{align}

Let $x_{a:i}(t)$ be the probability that a random walker starting from
node $a \in V$ is at node $i \in V$ at time $t$, where $\sum_{i \in V}
x_{a:i}(t) = 1$.  If Eq.~\eqref{eq:def_trans_prop} is used, then
$x_{a:i}(t+1)$ is
\begin{align}
  x_{a:i}(t+1) &= \sum_{j \in \partial i} x_{a:j}(t)\,p_{j \rightarrow i}.
  \label{eq:evol_existing_prob}
\end{align}
Using column vector $\vbf{x}_a(t) = (x_{a:i}(t))_{i \in V}$,
Eq.~\eqref{eq:evol_existing_prob} for all nodes $\forall i \in V$ can
be written simultaneously as
\begin{align}
  \vbf{x}_a(t + 1) &= \vbf{A}\,\vbf{D}^{-1}\,\vbf{x}_a(t),
  \label{eq:RW_from_a}
\end{align}
where $\vbf{D}$ and $\vbf{A}$ are the degree and adjacency matrices
defined as
\begin{align}
  \vbf{D} &:= \diag{d_i}_{i \in V},\\
  \vbf{A} &:= \left\{
  \begin{array}{ll}
    w_{ij} & \mathrm{if}\ \ (i, j) \in E\\
    0 & \mathrm{otherwise}
  \end{array}
  \right.,
\end{align}
respectively. $\vbf{A}\,\vbf{D}^{-1}$ is the matrix whose $(i, j)$th
element is the transition probability $p_{j \rightarrow
  i}$. Equation~\eqref{eq:RW_from_a} describes the behavior of the
random walk. Since $\vbf{A}\,\vbf{D}^{-1}$ is an asymmetric matrix,
Eq.~\eqref{eq:RW_from_a} is not easy to handle analytically using
linear algebra. Consequently, we modify Eq.~\eqref{eq:RW_from_a} to
\begin{align}
  \vbf{D}^{-1/2}\,\vbf{x}_a(t+1) &= \vbf{D}^{-1/2}\,\vbf{A}\,\vbf{D}^{-1/2}\,\vbf{D}^{-1/2}\,\vbf{x}_a(t) \nonumber\\
  &= \vbf{W}\,\vbf{D}^{-1/2}\,\vbf{x}_a(t)\nonumber\\
  \vbf{\hat{x}}_a(t+1) &= \vbf{W}\,\vbf{\hat{x}}_a(t),
  \label{eq:dyna_RW_vec}
\end{align}
where $\vbf{W} = \vbf{D}^{-1/2}\,\vbf{A}\,\vbf{D}^{-1/2}$ and
$\vbf{\hat{x}_a}(t) = \vbf{D}^{-1/2}\,\vbf{x}_a(t)$.  Since $\vbf{W}$
is a symmetric matrix, Eq.~\eqref{eq:dyna_RW_vec} is easier to handle
than Eq.~\eqref{eq:RW_from_a}.  In general, in spectral graph theory,
the behavior of an analysis target is expressed in terms of a matrix
such as Eq.~\eqref{eq:dyna_RW_vec}, and the characteristics of the
target are analyzed using the eigenvalues and eigenvectors of the
matrix on the basis of linear algebra.

$\vbf{W}$ can always be diagonalized using the orthogonal matrix
$\vbf{Q}$ that satisfies $\tmp{\vbf{Q}} = \vbf{Q}^{-1}$.  Let
$\lambda_k$ be the $k$th largest eigenvalue of $\vbf{W}$.  Note that
the maximum eigenvalue $\lambda_1$ is always $1$.  In this paper, we
assume that $G$ is connected and not a bipartite graph.  In this case,
the eigenvalues $\lambda_k$ for $ 2 \le k \le n$ satisfy
\begin{align}
  -1 < \lambda_n < \cdots < \lambda_2 < 1.
\end{align}
We let $\vbf{q}_k$ be the eigenvector for eigenvalue $\lambda_k$ by,
with the consequence that $\vbf{Q}$ is $\vbf{Q}= (\vbf{q}_k)_{1 \le k
  \le n}$.  Since $\vbf{Q}$ is an orthogonal matrix, $\vbf{q}_k$ and
$\vbf{q}_l$ for $1 \le k, l \le n$ satisfy
\begin{align}
  \tmp{\vbf{q}_k}\,\vbf{q}_l = \left\{
  \begin{array}{ll}
    1 & \mathrm{if} \ \ k = l \\
    0 & \mathrm{oterwise}
  \end{array}
  \right.
\end{align}
In particular, the maximum eigenvector $\vbf{q}_1$ is 
\begin{align} 
  \vbf{q}_1 = \frac{1}{\sqrt{s_1}}\tmp{(\sqrt{d_1}, \sqrt{d_2}, \cdots, \sqrt{d_n})},
\end{align}
where $s_1 = \sum_{i \in V} d_i$. $s_1$ is related to a statistic of
the graph structure of $G$ and can be written by $s_1 = n\,d_{\rm
  avg}$, where $d_{\rm avg}$ is the average weighted degree.

In~\cite{Lovasz96:RW}, Lov{\'a}sz analyzed a single random walk on
graph $G$ on the basis of spectral graph theory.  Solving
Eq.~\eqref{eq:dyna_RW_vec} in~\cite{Lovasz96:RW} results in the
  probability $x_{a:i}(t)$ being
\begin{align}
  x_{a:i}(t) &= \frac{\sqrt{d_i}}{\sqrt{d_a}} \sum_{k = 1}^{n} q_k(a)\,q_k(i)\, \lambda_k^t.
  \label{eq:sol_x}
\end{align}
According to this equation, $x_{a:i}(t)$ can be calculated using
eigenvalues $\lambda_k$ and eigenvectors $\vbf{q}_k$.  A closed-form
formula using eigenvalues and eigenvectors such as
Eq.~\eqref{eq:sol_x} is referred to as a {\it spectral formula}.

Let $x_{a:i}^*$ be the limit value of $x_{a:i}(t)$ for $t
\rightarrow\infty$.  From Eq.~\eqref{eq:sol_x}, $x_{a:i}^*$ can be
derived as
\begin{align}
  x_{a:i}^* &= \lim_{t \rightarrow \infty} x_{a:i}(t) \nonumber\\
  &= \lim_{t \rightarrow \infty} \left[\frac{\sqrt{d_i}}{\sqrt{d_a}} q_1(i)\,q_1(s) + \frac{\sqrt{d_i}}{\sqrt{d_a}} \sum_{k = 2}^{n} q_k(a)\,q_k(i)\,\lambda_k^t \right]\nonumber\\
  &= \frac{\sqrt{d_i}}{\sqrt{d_a}} q_1(a)\,q_1(i) = \frac{d_i}{s_1}.
  \label{eq:eq_x}
\end{align}
In this derivation process, we used $|\lambda_k| < 1$ for $k \ge2$.
According to Eq.~\eqref{eq:eq_x}, $x_{a:i}(t)$ is roughly proportional
to the weighted degree $d_i$ if sufficient time has elapsed since the
random walker started.

The analysis in~\cite{Lovasz96:RW} derived the expected first hitting
time $\mu_{a:i}$, the expected time it takes for a random walker
starting from node $a$ to arrive at node $i$.  From
Eq.~\eqref{eq:sol_x}, the spectral formula of expected first hitting
time $\mu_{a:i}$ is derived as
\begin{align}
  \mu_{a:i} &= s_1 \sum_{k=2}^{n} \frac{1}{1-\lambda_k} \left(\frac{q_k^2(i)}{d_i} - \frac{q_k(a)\,q_k(i)}{\sqrt{d_a\,d_i}}\right).
  \label{eq:first_arrival_time}
\end{align}

In~\cite{Von11:Hitting}, the effect of the graph structure on
$\mu_{a:i}$ was clarified using the spectral formula of the expected
first hitting time $\mu_{a:i}$.  According to~\cite{Von11:Hitting},
$\mu_{a:i}$ satisfies
\begin{align}
  \left| \frac{1}{s_1}\mu_{a:i} - \frac{1}{d_i} \right|
  \le \frac{2\,w_\mathrm{max}}{d^2_\mathrm{min}} \left( \frac{1}{1-\lambda_2} + 1\right),
  \label{eq:diff_mu_and_degree}
\end{align}
where $w_\mathrm{max}$ and $d_\mathrm{min}$ are the maximum of link
weights and the minimum of weighted degrees, respectively.  If the
right-hand side of Eq.~\eqref{eq:diff_mu_and_degree} is sufficiently
small, the expected first hitting time $\mu_{a:i}$ is approximated by
\begin{align}
  \mu_{a:i} \approx \frac{s_1}{d_i}.
  \label{eq:approx_mu}
\end{align}
In this case, $\mu_{a:i}$ is almost dominated by $s_1/d_i$, with the
consequence that $s_1/d_i$ can be expected to be the principal
component of $\mu_{a:i}$.  According to Eq.~\eqref{eq:approx_mu},
$\mu_{a:i}$ is roughly proportional to $s_1$, which is a statistic of
the graph structure.  In other words, $\mu_{a:i}$ corresponds to the
search time of node $i$ using the random walk.  Therefore,
Eq.~\eqref{eq:approx_mu} is also important for understanding the
characteristics of the search algorithm using a random walk.

\section{Analysis}
\label{sec:analysis}

In this section, we analyze the expected first meeting time
$\mu_{a,b}$ of two random walkers starting from node $a \in V$ and $b
\in V$ in graph $G$ on the basis of spectral graph theory.  We first
derive the spectral formula of $\mu_{a,b}$.  Then, we clarify the
principal component of $\mu_{a,b}$ using the derived spectral
formula.  Finally, we reveal the effect of the graph structure on
$\mu_{a,b}$ on the basis of the clarified principal component.

Our analysis results are important also for understanding the first
meeting time of $n_{\rm RW}$ random walkers, where $n_{\rm RW} > 2$,
because it is strongly affected by the first meeting time of two
random walkers.  To attain an efficient meeting of $n_{\rm RW}$ random
walkers, two of the $n_{\rm RW}$ random walkers must move together
after meeting at the same node.  In this case, the first meeting time
of $n_{\rm RW}$ random walkers is obtained as the sum of the first
meeting times of two random walkers.  Consequently, the characteristics
of the first meeting time for $n_{\rm RW}$ random walkers can be
expected to be strongly associated with that of two random walkers.

\subsection{Spectral Formula of Expected First Meeting Time $\mu_{a,b}$}

We derive the spectral formula of the expected first meeting time
$\mu_{a,b}$ using the same method as is used to derive that of the
expected first hitting time $\mu_{a:i}$
in~\cite{Lovasz96:RW}.  In~\cite{Lovasz96:RW}, the spectral formula was
derived using the generating function of the existing probability
$x_{a:i}(t)$.  In general, the generating function $F(z)$ of the
probability $f(t)$ is
\begin{align}
  F(z) := \sum_{t = 0}^{\infty} f(t) z^t.
  \label{eq:def_generation}
\end{align}
Using the generating function $F(z)$, the expectation $\ex{t}$ with
the probability $f(t)$ is
\begin{align}
  \ex{t} = \sum_{t = 1}^{\infty} t\,f(t) = \left. \frac{\diff F(z)}{\diff z} \right|_{z = 1}.
  \label{eq:exp_from_gen_func}
\end{align}
Importantly, even if we do not know the clused-form formula of the
probability $f(t)$, we can still derive the expectation $\ex{t}$ using
the generating function $F(z)$ on the basis of the above equation.  We
first obtain the generating function of the first meeting probability.
Without the value of the first meeting probability, we then derive the
spectral formula of the expected first meeting time $\mu_{a,b}$ by
substituting the generating function obtained into
Eq.~\eqref{eq:exp_from_gen_func}.

Let $r_{a,b:c}(t)$ be the probability that two random walkers first
meet at node $c$ at time $t$.  Since two random walkers can meet at any
node, the first meeting probability $r_{a,b:*}(t)$ is
\begin{align}
  r_{a,b:*}(t) = \sum_{c \in V} r_{a,b:c}(t).
\end{align}
In $r_{a,b:*}(t)$, the symbol $*$ designates any node in $V$.  Deriving
the spectral formula of the expected first meeting time $\mu_{a,b}$
using Eq.~\eqref{eq:exp_from_gen_func} requires the generating
function $R_{a,b:*}(z)$ of $r_{a,b:*}(t)$.

The probabilities that the two random walkers are at node $i$ at time
$t$ are $x_{a:i}(t)$ and $x_{b:i}(t)$, respectively.  Since the
spectral formulas of $x_{a:i}(t)$ and $x_{b:i}(t)$ are given by
Eq.~\eqref{eq:sol_x}, the generating functions of $x_{a:i}(t)$ and
$x_{b:i}(t)$ can be derived using Eq.~\eqref{eq:def_generation}.
However, since it is not easy to obtain the spectral formula of
$r_{a,b:*}(t)$, we obtain the generating function $R_{a,b:*}(z)$ of
$r_{a,b:*}(t)$ from those of $x_{a:i}(t)$ and $x_{b:i}(t)$ and then
derive the spectral formula of the expected first meeting time
$\mu_{a,b}$ using Eq.~\eqref{eq:exp_from_gen_func}.

With the aim of obtaining the generating function $R_{a,b:*}(z)$ of
the first meeting probability $r_{a,b:*}(t)$, we discuss the
relationship between $r_{a,b:*}(t)$, $x_{a:i}(t)$ and
$x_{b:i}(t)$.  Let $x_{a,b:c}(t)$ be the probability that the two
random walker meet at the same node $c \in V$ at time $t$.  The meeting
probability $x_{a, b:*}(t)$ at any node is
\begin{align}
  x_{a, b:*}(t) = \sum_{c \in V} x_{a,b:c}(t).
\end{align}
Since each random walker moves independently, $x_{a, b:*}(t)$ is 
\begin{align}
  x_{a,b:*}(t) = \sum_{c \in V} x_{a,b:c}(t) = \sum_{c \in V} x_{a:c}(t)\,x_{b:c}(t).
  \label{eq:joint_x1}
\end{align}
$x_{a,b:*}(t)$ includes both the first meeting probability
$r_{a,b:*}(t)$ and also the probability of the second and subsequent
meetings.  Hence, as shown in
Fig.~\ref{fig:state_trans_for_first_meeting}, we divide the transition
of the two random walks from time $0$ to time $t$ into two
transitions, (a)~the transition until they first meet at time $s$, and
(b)~the rest transition.  The probability for the former transition is
the first meeting probability $r_{a,b:*}(t)$.  The probability for the
latter transition is the probability that the two random walkers
starting from same node $c' \in V$ at time $s$ meet again at the same
node $c \in V$ at time $t$.  Since node $c'$ and node $c$ can be any
node, we denote such a probability by $x_{*',*': *}(t-s)$.  With these
probabilities, $x_{a,b:*}(t)$ is
\begin{align}
  x_{a,b:*}(t) = \sum_{s=0}^{t} r_{a,b:*'}(s) \, x_{*', *': *}(t-s).
  \label{eq:conv_r}
\end{align}
Using the probability $x_{c',c':c}(t)$ that the two random walkers
starting at node $c' \in V$ at time $0$ meet again at node $c \in V$
at time $t$, we set $x_{*', *': *}(t)$ as
\begin{align}
  x_{*',*':*}(t) = \sum_{c' \in V} \frac{d_{c'}^2}{s_2} \sum_{c \in V} x_{c',c':c}(t),
  \label{eq:meeting_prob}
\end{align}
where $s_2 = \sum_{i \in V} d_i^2$.  The reason that $x_{*',*':*}(t)$
is not set as a simple sum of values of $x_{c',c':c}(t)$ in
Eq.~\eqref{eq:meeting_prob} is as follows.  According to
Eq.~\eqref{eq:eq_x}, the probability $x_{a:i}^*$ in the steady state
is proportional to the weighted degree $d_i$ of node $i$.  Therefore,
the probability of the first meeting of the two random walkers at node
$c'$ can be expected to be proportional to $d_{c'}^2$.  Consequently,
in the sum of Eq.~\eqref{eq:meeting_prob}, $x_{c',c':c}(t)$ is
weighted by $d_{c'}^2/s_2$.  In Sect.~\ref{sec:example}, the validity
of Eq.~\eqref{eq:meeting_prob} will be confirmed through numerical
examples.

\insertPDFfig[.49]{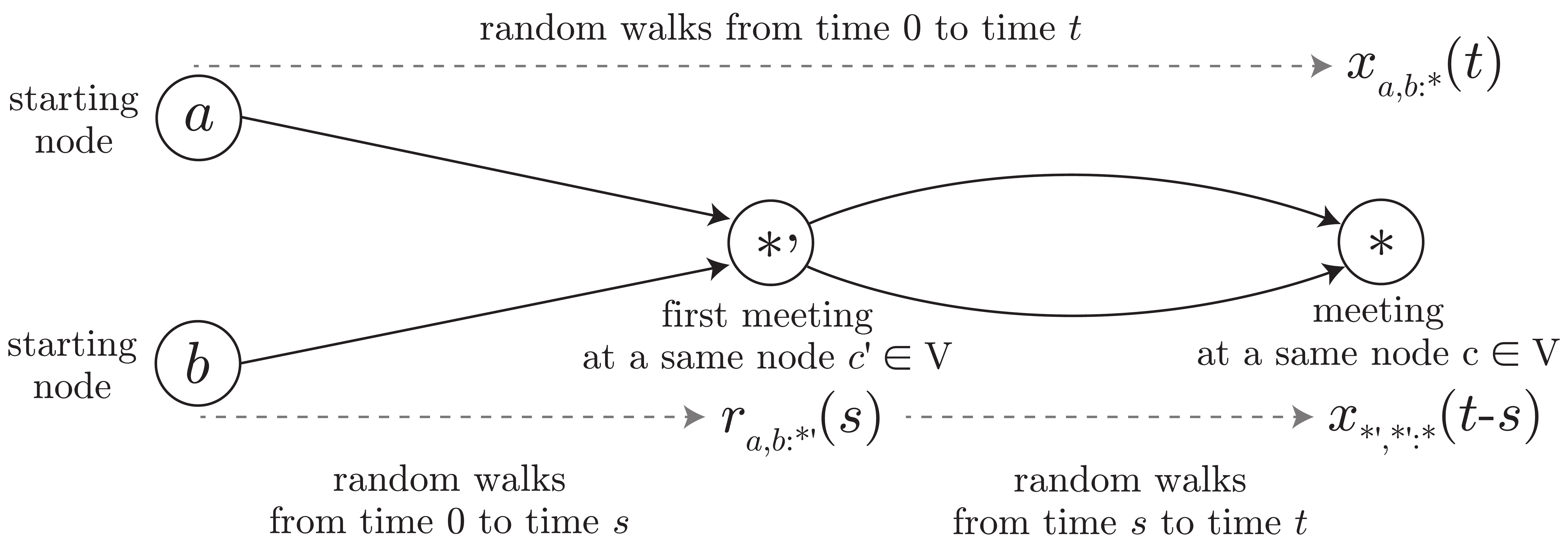}{Random walks starting from nodes $a$ and $b$ from time $0$ to time $t$}

The right-hand side of Eq.~\eqref{eq:conv_r} is a convolutional sum,
with the consequence that the generating function $X_{a,b:*}(z)$ of
$x_{a,b:*}(t)$ is
\begin{align}
  X_{a,b:*}(z) = R_{a,b:*'}(z) \, X_{*',*':*}(z).
  \label{eq:gen_R2_relation}
\end{align}
From this equation, the generating function $R_{a,b:*'}(z)$ of the
first meeting probability $r_{a,b:*}$ is
\begin{align}
  R_{a,b:*'}(z) = \frac{X_{a,b:*}(z)}{X_{*',*':*}(z)}.
  \label{eq:gen_R}
\end{align}
Substituting the spectral formulas of $x_{a:i}(t)$ and $x_{b:i}(t)$
given by Eq.~\eqref{eq:sol_x} into Eq.~\eqref{eq:def_generation}
yields the following spectral formula for $X_{a,b:*}(z)$:
\begin{align}
  & X_{a,b:*}(z) = \sum_{c \in V} \sum_{t=0}^{\infty} x_{a:c}(t)\,x_{b:c}(t)\,z^t \nonumber\\
  & \hspace{0.1cm} = \sum_{c \in V}\frac{d_c}{\sqrt{d_a d_b}}\!\sum_{k=1}^n \!\sum_{k'=1}^n \!q_k(a)q_k(c)q_{k'}(b)q_{k'}(c) \!\sum_{t=0}^{\infty} \!\left(\lambda_k \lambda_{k'} z\right)^t \nonumber\\
  & \hspace{0.1cm} = \sum_{c \in V}\frac{d_c}{\sqrt{d_a\,d_b}} \sum_{k=1}^n \sum_{k'=1}^n \frac{q_k(a)\,q_k(c)\,q_{k'}(b)\,q_{k'}(c)}{1-\lambda_k\,\lambda_{k'}\,z}.
  \label{eq:gen_Xab}
\end{align}
$z$ takes a value within the range of $|\lambda_k\,\lambda_{k'}\,z| <
1$ with the result that the sum of the infinite geometric series
converges.  From Eq.~\eqref{eq:meeting_prob}, the generating function
$X_{*',*':*}(z)$ is
\begin{align}
  X_{*',*':*}(z) = \sum_{c' \in V} \frac{d_{c'}^2}{s_2} \sum_{c \in V} X_{c',c':c}(z).
  \label{eq:gen_Xcc}
\end{align}

Substituting the generating function $R_{a,b:*'}(z)$ into
Eq.~\eqref{eq:exp_from_gen_func} results in the expected first meeting
time $\mu_{a,b}$ being
\begin{align}
  \mu_{a,b} = \sum_{s=0}^{\infty} s\,r_{a,b:*'}(s) = \left. \frac{\diff R_{a,b:*'}(z)}{\diff z} \right|_{z=1}.
  \label{eq:mu_gen}
\end{align}
In this equation, $\diff R_{a,b:*'}(z)/\diff z$ is
\begin{align}
  &\frac{\diff R_{a,b:*'}(z)}{\diff z} = \frac{\diff }{\diff z}\left(\frac{X_{a,b:*}(z)}{X_{*',*':*}(z)} \right) \nonumber\\
  &\hspace{1cm} = \frac{\dfrac{\diff X_{a,b:*}(z)}{\diff z} \, X_{*',*':*}(z) - X_{a,b:*}(z)\,\dfrac{\diff X_{*',*':*}(z)}{\diff z}}{X_{*',*':*}^2(z)} \nonumber\\
  &\hspace{1cm} = \frac{A(z) - B(z)}{C(z)},
  \label{eq:gen_R2_two_diff}
\end{align}
where
\begin{align}
  A(z) &= \frac{\diff X_{a,b:*}(z)}{\diff z} \, X_{*',*':*}(z), \label{eq:A}\\
  B(z) &= X_{a,b:*}(z)\,\frac{\diff X_{*',*':*}(z)}{\diff z}, \label{eq:B} \\
  C(z) &= X_{*',*':*}^2(z). \label{eq:C}
\end{align}
According to Eqs.~\eqref{eq:gen_Xab} and \eqref{eq:gen_Xcc}, $A(z)$,
$B(z)$, and $C(z)$ can be written as polynomials for $(1-z)$ because
of $1-\lambda_1 \lambda_1 z = 1-z$.  Hence, we also obtain $A(z)$ as
\begin{align}
  A(z) = \frac{A_3(z)}{(1-z)^3} + \frac{A_2(z)}{(1-z)^2} + \frac{A_1(z)}{1-z} + A_0(z).
\end{align}
Substituting Eqs.~\eqref{eq:gen_Xab} and \eqref{eq:gen_Xcc} into the
right-hand side of Eq.~\eqref{eq:A} yields $A_3(z)$ and $A_2(z)$ as
\begin{align}
  A_3(z) \!\!&=\!\! \left[ \sum_{c \in V} \frac{d_c}{\sqrt{d_a d_b}} q_1(a)\,q_1^2(c)\,q_1(b) \right]\nonumber\\
  &\ \ \ \ \ \ \ \ \ \ \ \ \ \ \left[ \sum_{c \in V} \sum_{c' \in V} \frac{d_{c'}\,d_c}{s_2} q_1^2(c)\,q_1^2(c') \right], \\
  A_2(z) \!\!&=\!\! \left[ \sum_{c \in V} \frac{d_c}{\sqrt{d_a d_b}} q_1(a)\,q_1^2(c)\,q_1(b) \right] \nonumber\\
  &\ \ \left[ \sum_{c \in V} \sum_{c' \in V} \!\frac{d_{c'} d_c}{s_2} \!\!\!\!\sum_{\substack{1 \le k,k' \le n\\ (k, k') \neq (1,1)}} \!\!\!\!\frac{q_k(c)q_k(c')q_{k'}(c)q_{k'}(c')}{1-\lambda_k\lambda_{k'}\,z} \right] \nonumber\\
  &\!\!\!=\!\! \frac{1}{s_1^2} \!\!\left[\sum_{c \in V} \sum_{c' \in V} d_{c'} d_c \!\!\!\!\sum_{\substack{1 \le k,k' \le n\\ (k, k') \neq (1,1)}} \!\!\!\!\!\frac{q_k(c)q_k(c')q_{k'}(c)q_{k'}(c')}{1-\lambda_k\lambda_{k'} z} \right]\!\!.
\end{align}
We do not provide $A_1(z)$ and $A_0(z)$ in this paper because
$A_1(z)/(1-z)$ and $A_0(z)$ disappear when deriving $\mu_{a,b}$ using
Eq.~\eqref{eq:mu_gen}.  Similarly, $B(z)$ is
\begin{align}
  B(z) = \frac{B_3(z)}{(1-z)^3} + \frac{B_2(z)}{(1-z)^2} + \frac{B_1(z)}{1-z} + B_0(z).
\end{align}
Substituting Eqs.~\eqref{eq:gen_Xab} and \eqref{eq:gen_Xcc} into the
right-hand side of Eq.~\eqref{eq:B} yields $B_3(z)$ and $B_2(z)$ as
\begin{align}
  B_3(z) \!\!&=\!\! \left[ \sum_{c \in V} \frac{d_c}{\sqrt{d_a d_b}} q_1(a)\,q_1^2(c)\,q_1(b) \right]\nonumber\\
  &\ \ \ \ \ \ \ \ \ \ \ \ \ \ \left[ \sum_{c \in V} \sum_{c' \in V} \frac{d_{c'}\,d_c}{s_2} q_1^2(c)\,q_1^2(c') \right], \\
  B_2(z) \!\!&=\!\! \left[ \sum_{c \in V} \sum_{\substack{1 \le k,k' \le n\\ (k, k')\neq (1,1)}} \frac{d_c}{\sqrt{d_a\,d_b}} \frac{q_k(a)\,q_k(c)\,q_{k'}(b)\,q_{k'}(c)}{1-\lambda_k\lambda_{k'}\,z} \right] \nonumber\\
  &\ \ \ \ \ \ \ \ \ \ \ \ \ \ \ \ \ \ \ \ \ \ \ \ \ \ \ \ \ \ \ \ \ \left[ \sum_{c \in V} \sum_{c' \in V} \frac{d_{c'}\,d_c}{s_2} q_1^2(c)\,q_1^2(c') \right] \nonumber\\
  &=\!\! \frac{1}{s_1^2}\!\! \left[\sum_{c \in V} \!\frac{d_c}{\sqrt{d_a\,d_b}}\!\!\!\!\sum_{\substack{1\le k,k' \le 1\\ (k, k') \neq (1,1)}} \!\!\!\! \frac{q_k(a)\,q_k(c)\,q_{k'}(b)\,q_{k'}(c)}{1-\lambda_k\lambda_{k'}\,z} \right].
\end{align}
Using these equations, we have found that $A_3(z) =
B_3(z)$.  Consequently, $A(z) - B(z)$ in the numerator of
Eq.~\eqref{eq:gen_R2_two_diff} does not contain the term $(1-z)^{-3}$,
with the result that the term $(1-z)^{-2}$ becomes the highest-order
term in the polynomials for $(1-z)$ in $A(z) - B(z)$.  Thus, $C(z)$ is
\begin{align}
  C(z) = \frac{C_2(z)}{(1-z)^2} + \frac{C_1(z)}{1-z} + C_0(z).
\end{align}
Solving this equation in the same manner yields $C_2(z)$ as
\begin{align}
  C_2(z) &= \left[\sum_{c \in V} \sum_{c' \in V} \frac{d_{c'}\,d_c}{s_2} q_1^2(c)\,q_1^2(c') \right]^2 \nonumber\\
  &= \left[ \frac{1}{s_1^2} \sum_{c \in V} d_c^2 \right]^2 = \frac{(s_2)^2}{s_1^4}.
\end{align}

Since $(A(z) - B(z))/C(z)$ is an indeterminate form at $z = 1$, we
discuss $\lim_{z \rightarrow 1} (A(z) - B(z))/C(z)$ to derive the
spectral formula of the expected first meeting time $\mu_{a,b}$ using
Eqs.~\eqref{eq:mu_gen} and \eqref{eq:gen_R2_two_diff}.  As the limit of
$z \rightarrow 1$, $\mu_{a,b}$ is
\begin{align}
  \mu_{a,b} &= \lim_{z \rightarrow 1} \frac{A(z) - B(z)}{C(z)} = \lim_{z \rightarrow 1} \frac{(1-z)^2\!\left(A(z) \!-\! B(z)\right)}{(1-z)^2\,C(z)} \nonumber\\
  &= \frac{A_2(1) - B_2(1)}{C_2(1)} \nonumber\\
  &= \frac{s_1^2}{(s_2)^2} \!\!\left[\sum_{c \in V} \sum_{c' \in V} \!d_{c'}d_c \!\!\!\!\!\!\!\!\sum_{\substack{1 \le k,k' \le n\\ (k, k') \neq (1,1)}} \!\!\!\!\!\!\!\!\frac{q_k(c)q_k(c')q_{k'}(c)q_{k'}(c')}{1\!-\!\lambda_k\lambda_{k'}} \right. \nonumber\\
    \!\!& \left. \ \ \ \ - \sum_{c \in V} \frac{d_c}{\sqrt{d_a d_b}} \!\!\!\!\sum_{\substack{1\le k,k' \le 1\\ (k, k') \neq (1,1)}} \!\!\!\!\frac{q_k(a)\,q_k(c)\,q_{k'}(b)\,q_{k'}(c)}{1\!-\!\lambda_k\lambda_{k'}} \right] \nonumber\\
  &= \frac{1}{(s_2)^2} \sum_{c \in V} d_c^2 \sum_{c' \in V} d_{c'}^2 \Biggl[ \Biggr. \nonumber\\
    s_1 & \sum_{k=2}^{n} \frac{1}{1-\lambda_k} \!\left[\frac{2q_k(c)q_k(c')}{\sqrt{d_c d_{c'}}} \!-\! \frac{q_k(c)}{\sqrt{d_c}}\!\left(\frac{q_k(a)}{\sqrt{d_a}} \!+\! \frac{q_k(b)}{\sqrt{d_b}} \right)\!\right] \nonumber\\
    + s_1^2 & \Biggl.\!\!\!\!\sum_{\substack{1\le k,k'\le n\\(k,k')\neq (1,1)}}\!\!\!\! \frac{q_k(c) q_{k'}(c)}{(1\!-\!\lambda_k\lambda_{k'}) d_c} \!\!\left(\! \frac{q_k(c') q_{k'}(c')}{d_{c'}} \!-\! \frac{q_k(a) q_{k'}(b)}{\sqrt{d_a\,d_b}}\!\right)\!\Biggr]. \nonumber\\
  \label{eq:avg_first_meeting_time_pre}
\end{align}
Since this equation is expressed by the eigenvalues and eigenvectors
of $\vbf{W}$, it is the spectral formula of $\mu_{a,b}$.

Equation~\eqref{eq:avg_first_meeting_time_pre} appears to be
complicated, but if we use the expected first meeting time
$\mu_{a,b:c}$, the time until the two random walkers first meet at
node $c \in V$, then $\mu_{a,b}$ is
\begin{align}
  \mu_{a,b} &= \frac{1}{s_2} \sum_{c \in V} d_c^2 \,\mu_{a,b:c} - \frac{1}{s_2^2} \sum_{c \in V} d_c^2 \sum_{c' \in V} d_{c'}^2 \,\mu_{c',c':c}, \nonumber\\
  \label{eq:avg_first_meeting_time}
\end{align}
where the spectral formula of $\mu_{a,b:c}$ is 
\begin{align}
  \mu_{a,b:c} &= \mu_{a:c} + \mu_{b:c} \nonumber\\
  + s_1^2 \!\!\!\!&\!\!\!\! \sum_{\substack{1\le k,k'\le n\\(k,k')\neq (1,1)}}\!\!\!\! \frac{q_k(c) q_{k'}(c)}{(1\!-\!\lambda_k\lambda_{k'}) d_c}\!\! \left( \!\frac{q_k(c) q_{k'}(c)}{d_c} \!-\! \frac{q_k(a) q_{k'}(b)}{\sqrt{d_a\,d_b}}\!\right). \nonumber\\
  \label{eq:avg_first_meet_time_at_c}
\end{align}

\subsection{Principal Component of the Expected First Meeting Time $\mu_{a,b}$}

We examine the principal component of $\mu_{a,b}$ with the spectral
formula of the expected first meeting time $\mu_{a,b}$ and reveal
mathematically the effect of the graph structure on the expected first
meeting time $\mu_{a,b}$ on the basis of the clarified principal
component.  We use the method for examining the first hitting time
$\mu_{a:i}$ used in~\cite{Von11:Hitting} to derive the principal
component of $\mu_{a,b}$.

First, we introduce
\begin{align}
  \hvbf{N} &:= \vbf{I} \otimes \vbf{I} - \vbf{W} \otimes \vbf{W} = \hvbf{I} - \hvbf{W},
\end{align}
where $\vbf{I}$ is the $n \times n$ unit matrix and $\otimes$ is the
Kronecker product.  According to the definition of the Kronecker
product, $\hvbf{I}$ and $\hvbf{W}$ are $n^2 \times n^2$ matrices.  Let
$\hvbf{N}^\dagger$ be the pseudo-inverse matrix of $\hvbf{N}$ with the
result that $\hvbf{N} \hvbf{N}^\dagger \hvbf{N} = \hvbf{N}$,
\begin{align}
  \hvbf{N}^\dagger = \sum_{\substack{1\le k,k'\le n\\(k,k')\neq (1,1)}} \frac{\hvbf{q}_{kk'}\,\tmp{\hvbf{q}_{kk'}}}{1-\lambda_k\lambda_{k'}},
  \label{eq:inv_N}
\end{align}
where $\hvbf{q}_{kk'}$ is the following column vector with $n^2$
elements:
\begin{align}
 \hvbf{q}_{kk'} &:= \vbf{q}_{k} \otimes \vbf{q}_{k'}.
\end{align}
Substituting $\hvbf{N}^\dagger$ into
Eq.~\eqref{eq:avg_first_meeting_time_pre} yields the following as the
expected first meeting time $\mu_{a,b}$:
\begin{align}
  \mu_{a,b} &= \frac{s_1^2}{s_2^2} \sum_{c \in V} \sum_{c' = 1}^n d_c^2 \, d_{c'}^2 \tmp{\hvbf{u}_{cc}} \hvbf{N}^\dagger (\hvbf{u}_{c'c'} - \hvbf{u}_{ab}).
  \label{eq:avg_first_meeting_time_N1}
\end{align}
In this equation, $\hvbf{u}_{ab}$ is 
\begin{align}
 \hvbf{u}_{ab} &:= \vbf{u}_{a} \otimes \vbf{u}_{b},
\end{align}
where $\vbf{u}_{a}$ is the column vector whose $i$th element
$\vbf{u}_{a}(i)$ is
\begin{align}
  \vbf{u}_{a}(i) &= \left\{
  \begin{array}{cc}
    \ds \frac{1}{\sqrt{d_a}} & \mathrm{if}\ \ \ i = a\\
    0 & \mathrm{otherwise}
  \end{array}
  \right..
\end{align}
The pseudo-inverse matrix $\hvbf{N}^\dagger$ of $\hvbf{N}$ in
Eq.~\eqref{eq:inv_N} is also
\begin{align}
 \hvbf{N}^\dagger &= \sum_{\substack{1\le k,k'\le n\\(k,k')\neq (1,1)}} \frac{\hvbf{q}_{kk'}\,\tmp{\hvbf{q}_{kk'}}}{1-\lambda_k\lambda_{k'}}\nonumber\\
     &= \sum_{\substack{1\le k,k'\le n\\(k,k')\neq (1,1)}} \frac{(1-\lambda_k\lambda_{k'}+\lambda_k\lambda_{k'})\,\hvbf{q}_{kk'}\,\tmp{\hvbf{q}_{kk'}}}{1-\lambda_k\lambda_{k'}}\nonumber\\
     &= \hvbf{I} - \hvbf{q}_{11}\tmp{\hvbf{q}_{11}} + \hvbf{M},
 \label{eq:inv_N2}
\end{align}
where $\hvbf{M}$ is 
\begin{align}
 \hvbf{M} &= \sum_{\substack{1\le k,k'\le n\\(k,k')\neq (1,1)}} \frac{\lambda_k\,\lambda_{k'}\,\hvbf{q}_{kk'}\,\tmp{\hvbf{q}_{kk'}}}{1-\lambda_k\lambda_{k'}}\nonumber\\
   &= \sum_{\substack{1\le k,k'\le n\\(k,k')\neq (1,1)}} \sum_{s=1}^{\infty} \left(\lambda_k\,\lambda_{k'}\hvbf{q}_{kk'}\,\tmp{\hvbf{q}_{kk'}}\right)^s\nonumber\\
   &= \sum_{s=1}^{\infty} \left(\sum_{\substack{1\le k,k'\le n\\(k,k')\neq (1,1)}} \lambda_k\,\lambda_{k'}\hvbf{q}_{kk'}\,\tmp{\hvbf{q}_{kk'}}\right)^s\nonumber\\
   &= \sum_{s=1}^{\infty} \left(\sum_{k=1}^n\sum_{k'=1}^n \lambda_k\,\lambda_{k'}\,\hvbf{q}_{kk'}\,\tmp{\hvbf{q}_{kk'}} - \hvbf{q}_{11}\,\tmp{\hvbf{q}_{11}}\right)^s\nonumber\\
   &= \sum_{s=1}^{\infty} \left(\hvbf{W} - \hvbf{q}_{11}\,\tmp{\hvbf{q}_{11}}\right)^s \nonumber\\
   &= \hvbf{W} \!-\! \hvbf{q}_{11}\!\tmp{\hvbf{q}_{11}} \!+\! \left(\hvbf{W} \!-\! \hvbf{q}_{11}\!\tmp{\hvbf{q}_{11}}\right)^2 \sum_{s=0}^{\infty} \!\left(\hvbf{W} \!-\! \hvbf{q}_{11}\tmp{\hvbf{q}_{11}}\right)^s\nonumber\\
   &= \hvbf{W} \!-\! \hvbf{q}_{11}\tmp{\hvbf{q}_{11}} \!+\! \left(\hvbf{W} \!-\! \hvbf{q}_{11}\tmp{\hvbf{q}_{11}}\right)^2 \!\!\!\!\sum_{\substack{1\le k,k'\le n\\(k,k')\neq (1,1)}}\!\!\frac{\hvbf{q}_{kk'}\tmp{\hvbf{q}_{kk'}}}{1 \!-\! \lambda_k\lambda_{k'}}.
\end{align}
This derivation process involved the use of
\begin{align}
  \tmp{\hvbf{q}_{ij}}\,\hvbf{q}_{kl} &= \left\{
  \begin{array}{ll}
    1 & \mathrm{if} \ i = k \ \mathrm{and}\ j = l\\
    0 & \mathrm{otherwise}
  \end{array}
  \right.,\\
  (\hvbf{q}_{kk'}\,\tmp{\hvbf{q}_{kk'}})^s &= \hvbf{q}_{kk'}\,\tmp{\hvbf{q}_{kk'}}.
\end{align}
Substituting Eq.~\eqref{eq:inv_N2} into
Eq.~\eqref{eq:avg_first_meeting_time_N1} yields the following as
$\mu_{a,b}$:
\begin{align}
  \mu_{a,b} &= \frac{s_1^2}{s_2} + \frac{s_1^2}{s_2^2} \sum_{c \in V} \sum_{c' = 1}^n d_c^2 \, d_{c'}^2 \tmp{\hvbf{u}_{cc}} \hvbf{M} (\hvbf{u}_{c'c'} - \hvbf{u}_{ab}).
  \label{eq:avg_first_meeting_time_N2}
\end{align}
The following was used to obtain this equation: 
\begin{align}
  \tmp{\hvbf{u}_{cc}} &\left( \hvbf{N}^\dagger (\hvbf{u}_{c'c'} - \hvbf{u}_{ab}) \right) \nonumber\\
  &= \tmp{\hvbf{u}_{cc}} \left[\left(\hvbf{I} - \hvbf{q}_{11}\tmp{\hvbf{q}_{11}} + \hvbf{M}\right) (\hvbf{u}_{c'c'} - \hvbf{u}_{ab}) \right] \nonumber\\
  &= \left\{
  \begin{array}{ll}
    \ds\frac{1}{d_c^2} + \tmp{\hvbf{u}_{cc}} \hvbf{M} (\hvbf{u}_{c'c'} - \hvbf{u}_{ab}) & \mathrm{if}\ c = c' \\
    \ds\tmp{\hvbf{u}_{cc}} \hvbf{M} (\hvbf{u}_{c'c'} - \hvbf{u}_{ab}) & \mathrm{otherwise} \\
  \end{array}
  \right..
\end{align}
The first term on the right-hand side of
Eq.~\eqref{eq:avg_first_meeting_time_N2} corresponds to the principal
component of the expected first meeting time $\mu_{a,b}$.

To confirm that $s_1^2/s_2$ is the principal component of the expected
first meeting time $\mu_{a,b}$, we discuss
\begin{align}
  \left| \frac{\mu_{a,b}}{s_1^2} - \frac{1}{s_2} \right| = \frac{1}{s_2^2} \sum_{c \in V} \sum_{c' = 1}^n d_c^2 \, d_{c'}^2 \left|\tmp{\hvbf{u}_{cc}} \hvbf{M} (\hvbf{u}_{c'c'} - \hvbf{u}_{ab}) \right|.
  \label{eq:diff_meet_and_sd}
\end{align}
The right-hand side of this equation expresses the error between
$\mu_{a,b}$ and the principal component $s_1^2/s_2$.

We examine the upper bound on the right-hand side of
Eq.~\eqref{eq:diff_meet_and_sd} using
\begin{align}
  \frac{1}{1-\lambda_k\,\lambda_{k'}} & \le \frac{1}{1-\lambda_2},
\end{align}
for $2 \le k, k' \le n$.  Using the above equation, we obtain
\begin{align}
  & \left|\tmp{\hvbf{u}_{cc}} \hvbf{M} (\hvbf{u}_{c'c'} - \hvbf{u}_{ab}) \right|\nonumber\\
  & = \left|\tmp{\hvbf{u}_{cc}}\left(\hvbf{W} \!-\! \hvbf{q}_{11}\tmp{\hvbf{q}_{11}}\right) (\hvbf{u}_{c'c'} - \hvbf{u}_{ab}) \right|\nonumber\\
  & \ \ +\!\left|\tmp{\hvbf{u}_{cc}}\left(\hvbf{W} \!-\! \hvbf{q}_{11}\tmp{\hvbf{q}_{11}}\right)^2 \!\!\!\!\!\!\!\sum_{\substack{1\le k,k'\le n\\(k,k')\neq (1,1)}}\!\!\!\frac{\hvbf{q}_{kk'}\tmp{\hvbf{q}_{kk'}}}{1 \!-\! \lambda_k\lambda_{k'}} (\hvbf{u}_{c'c'} - \hvbf{u}_{ab}) \right|\nonumber\\
  & \le \left|\tmp{\hvbf{u}_{cc}}\left(\hvbf{W} \!-\! \hvbf{q}_{11}\tmp{\hvbf{q}_{11}}\right)\!(\hvbf{u}_{c'c'} \!-\! \hvbf{u}_{ab}) \right|\nonumber\\
  &\ \ +\! \frac{1}{1 \!-\! \lambda_2} \!\left|\tmp{\hvbf{u}_{cc}} \!\left(\!\hvbf{W} \!-\! \hvbf{q}_{11}\tmp{\hvbf{q}_{11}}\right)^{\!2}\!\!\left(\!\hvbf{I} \!-\! \hvbf{q}_{11}\tmp{\hvbf{q}_{11}}\right)\!(\hvbf{u}_{c'c'} \!-\! \hvbf{u}_{ab}) \right|\nonumber\\
  & = \left|\tmp{\hvbf{u}_{cc}}\hvbf{W}\!(\hvbf{u}_{c'c'} \!-\! \hvbf{u}_{ab}) \right| \!+\! \frac{1}{1 \!-\! \lambda_2} \left|\tmp{\hvbf{u}_{cc}} \hvbf{W}^2(\hvbf{u}_{c'c'} \!-\! \hvbf{u}_{ab}) \right|\nonumber\\
  &= \left|\tmp{\hvbf{u}_{cc}}\hvbf{W}\!(\hvbf{u}_{c'c'} \!-\! \hvbf{u}_{ab}) \right| \!+\! \frac{1}{1 \!-\! \lambda_2} \left|\hvbf{W}\hvbf{u}_{cc}\right| \!\left|\hvbf{W}\!(\hvbf{u}_{c'c'} \!-\! \hvbf{u}_{ab}) \right|\nonumber\\
  & \le \frac{2\,w_\mathrm{max}^2}{d_\mathrm{min}^4} + \frac{1}{1 \!-\! \lambda_2} \frac{w_\mathrm{max}}{d_\mathrm{min}^2}\frac{\sqrt{2}\,w_\mathrm{max}}{d_\mathrm{min}^2}\nonumber\\
  & \le \frac{2\,w_\mathrm{max}^2}{d_\mathrm{min}^4} \left(\frac{1}{1 \!-\! \lambda_2} + 1 \right).
  \label{eq:upper_umv}
\end{align}
The following was used in this derivation process:
\begin{align}
  \left|\tmp{\hvbf{u}_{ij}}\hvbf{W}\hvbf{u}_{kl} \right| &= \tmp{\vbf{u}}_i\vbf{W}\vbf{u}_k \otimes \tmp{\vbf{u}}_j\vbf{W}\vbf{u}_l \nonumber\\
  &= \frac{w_{ik}}{d_i\,d_k} \otimes \frac{w_{jl}}{d_j\,d_l} \le \frac{w_\mathrm{max}^2}{d_\mathrm{min}^4}, \\
  \left|\hvbf{W}\hvbf{u}_{ij} \right|^2 &= \tmp{\left(\hvbf{W}\hvbf{u}_{ij}\right)} \left(\hvbf{W}\hvbf{u}_{ij}\right) \nonumber\\
  &= \sum_{k \in \partial i}^n\sum_{l \in \partial j}^n \frac{w_{ki}^2\,w_{lj}^2}{d_k\,d_i^2\,d_l\,d_j^2} \nonumber\\
  &\le\frac{w_\mathrm{max}^2}{d_i^2\,d_j^2\,d_\mathrm{min}^2} \sum_{k \in \partial i}^n \sum_{l \in \partial j}^n w_{ki}\,w_{lj} \nonumber\\
  &= \frac{w_\mathrm{max}^2}{d_i\,d_j\,d_\mathrm{min}^2} \le \frac{w_\mathrm{max}^2}{d_\mathrm{min}^4}.
\end{align}

Substituting Eq.~\eqref{eq:upper_umv} into
Eq.~\eqref{eq:diff_meet_and_sd} yields the following as the upper
bound of the error between $\mu_{a,b}$ and the principal component
$s_1^2/s_2$:
\begin{align}
  \left| \frac{\mu_{a,b}}{s_1^2} - \frac{1}{s_2} \right| \le \frac{2\,w_\mathrm{max}^2}{d_\mathrm{min}^4} \left(\frac{1}{1 - \lambda_2} + 1 \right).
  \label{eq:err_in_meet}
\end{align}
According to this equation, the error can be expected to be small for
graph $G$, where $\lambda_2$ and $w_\mathrm{max}$ are small but
$d_\mathrm{min}$ is large.  In this case, the expected first meeting
time $\mu_{a,b}$ is approximated as
\begin{align}
  \mu_{a,b} \approx \frac{s_1^2}{s_2} = \frac{n}{1 + d_{\rm std}^2/d_{\rm avg}^2},
  \label{eq:approx_meet}
\end{align}
where $d_{\rm avg}$ and $d_{\rm std}$ are the average and standard
deviation of weighted degrees, respectively.

If the approximation formula~\eqref{eq:approx_meet} holds for the
expected first meeting time $\mu_{a,b}$, we derive the following
characteristics: (a)~$\mu_{a,b}$ is small when the coefficient of
variation $d_{\rm std}/d_{\rm avg}$ is large and (b)~$\mu_{a,b}$ does
not depend on the starting nodes $a$ and $b$.  The characteristic~(a)
is useful for understanding the effect of the graph structure on
$\mu_{a,b}$.

\section{Numerical Example}
\label{sec:example}

In this section, we confirm the validity of the spectral
formula~\eqref{eq:avg_first_meeting_time_pre} and the principal
component of the expected first meeting time $\mu_{a,b}$ revealed in
Sect.~\ref{sec:analysis}. We also examine the error in the
approximation formula~\eqref{eq:approx_meet} obtained when $\mu_{a,b}$
is replaced by its principal component.

\subsection{Setting}

In this subsection, we use BA~(Barab\'asi-Albert)
graphs~\cite{Barabasi99:Model} and ER~(Erd\"os-R\'enyi)
graphs~\cite{Erdos59:Random}.  The spectral
formula~\eqref{eq:avg_first_meeting_time_pre} and the approximation
formula~\eqref{eq:approx_meet} depend on the degree distribution of a
graph. Since the degree distribution of a BA graph is different from
that of an ER graph, these graphs are useful for clarifying the
effects of the degree distribution on these formulas.  Owing to space
limitations, we provide the results for unweighted graphs, where
$w_{ij} = 1$ is provided for all links $\forall (i, j) \in E$. In
unweighted graphs, the weighted degree $d_i$ of node $i$ corresponds
to the degree $m_i$, the number of links of node $i$.

The BA model~\cite{Barabasi99:Model} is a typical model for scale-free
random graphs.  A BA graph is generated using the following procedure.
First, a complete graph with $n_0$ nodes is created.  We assume that
$n_0 = m$ for the sake of simplicity.  Next, nodes are inserted one by
one until the number of nodes in the BA graph is equal to $n$.  When
adding the $t$th node~($t = m + 1, m + 2, \cdots, n$), $m_{\rm new}$
links are created from node $t$ to nodes $j \in \{1, 2, \cdots, t-1\}$
with the connection probability $p^\mathrm{BA}_j(t)$.  The connection
probability $p^\mathrm{BA}_j(t)$ is
\begin{align}
  p^\mathrm{BA}_j(t) = \frac{m_j(t)}{\sum_{l = 1}^{t-1} m_l(t)},
\end{align}
where $m_j(t)$ is the degree of node $j$ when the insertion of the
$t-1$th node is completed.  BA graphs have the power-law degree
distribution~(i.e, $\prob{m_i = m} \sim m^{-3}$).  If $G$ is
unweighted, then the average weighted degree $d_{\rm avg}$ is equal to
the average degree $k_{\rm avg}$.  Hence, the average weighted degree
$d_{\rm avg}$ of a BA graph is approximated as
\begin{align}
  d_{\rm avg} = \frac{m\,(m - 1) + 2\,m\,(n - m)}{n} \approx 2\,m,
  \label{eq:davg_ba}
\end{align}
where we assume that $n \gg m$.

In contrast, the ER model~\cite{Erdos59:Random} is a classical random
graph model.  An ER graph is generated through the following
procedure.  First, $n$ nodes are created.  Next, links are created
between any pair of nodes with probability $p^\mathrm{ER}$.  If the
graph is not connected, then the link creation process is begun again.
The average weighted degree $d_{\rm avg}$ of an ER graph is
\begin{align}
 d_{\rm avg} &= (n-1)\,p^\mathrm{ER}.
 \label{eq:davg_er}
\end{align}
The degree distribution $\prob{m_i = m}$ of an ER graph follows the
binomial distribution.  According to the difference between the power
law and the binomial distribution, the standard deviation $d_{\rm
  std}$ of weighted degrees in a BA graph is greater than that in an
ER graph.

To focus on the difference in the standard deviation $d_{\rm std}$ of
weighted degrees, we set $m$ and $p^\mathrm{ER}$ as
\begin{align}
 m &= \left\lfloor \frac{d_{\rm avg}}{2} \right\rfloor,\\
 p_{\rm ER} &= \frac{d_{\rm avg}}{n - 1},
\end{align}
with the result that the average weighted degree $d_{\rm avg}$ of an ER
graph and a BA graph are roughly equal.  The minimum weighted degree
$d_{\rm min}$ in both graphs also increases as $d_{\rm avg}$
increases.

To examine the validity and the error of the spectral
formula~\eqref{eq:avg_first_meeting_time_pre} and the approximation
formula~\eqref{eq:approx_meet}, we measure the average of the first
meeting times in simulation using the following procedure.
\begin{enumerate}
\item Generate a BA graph or an ER graph using the above procedures.
\item Put random walkers on nodes $a \in V$ and $b \in V$.
\item Move each random walker with the transition probability $p_{i
  \rightarrow j}$ in accordance with Eq.~\eqref{eq:def_trans_prop}.
\item Repeat step 3 until the two random walks meet at the same node.
\item Repeat step 2 through step 4 10,000 times to calculate the
  average of the first meeting times.
\end{enumerate}
We use the parameter configuration shown in Tab.~\ref{tab:param} as a
default parameter configuration.

\begin{table}[tb]
 \caption{Parameter configuration}
 \begin{center}
 \begin{tabular}{lc|c} 
  \hline
  Number of nodes & $n$ & 1,000\\
  Weight of link $(i, j)$ & $w_{ij}$ & 1\\
  Average weighted degree & $d_\mathrm{avg}$ & 6\\
  Random walker's starting node & $a$ & 1\\
  \hline
 \end{tabular}
 \label{tab:param}
 \end{center}
\end{table}

\subsection{Validity of the Spectral Formula for the Expected First Meeting Time $\mu_{a,b}$}

We confirm the validity of the spectral formula of the expected first
meeting time $\mu_{a,b}$ given by
Eq.~\eqref{eq:avg_first_meeting_time_pre}.

Figures~\ref{fig:N-avg_first_meettime_ba} and
\ref{fig:N-avg_first_meettime_er} show the first meeting times
obtained from the simulation and the analysis~(i.e., the spectral
formula~\eqref{eq:avg_first_meeting_time_pre}) for different settings
of the random walker's starting node $b$ in the BA and ER graphs,
respectively.  According to Figs.~\ref{fig:N-avg_first_meettime_ba} and
\ref{fig:N-avg_first_meettime_er}, the analysis results are almost the
same as the simulation results regardless of the choice of $n$ and
$b$.

\insertfig{N-avg_first_meettime_ba}{Expected first meeting time $\mu_{a, b}$ for different settings of the starting node $b$ of the random walker in BA graphs}
\insertfig{N-avg_first_meettime_er}{Expected first meeting time $\mu_{a, b}$ for different settings of the starting node $b$ of the random walker in ER graphs}

We then evaluate the error in the spectral
formula~\eqref{eq:avg_first_meeting_time_pre}.  In this evaluation, we
use the relative error $\epsilon_{a,b}$ of the expected first meeting
time $\mu_{a,b}$.  The relative error $\epsilon_{a,b}$ is defined as
\begin{align}
  \epsilon_{a,b} := \frac{\left|\mu_{a,b} - \mu_{a,b}^\mathrm{sim}\right|}{\mu_{a,b}^\mathrm{sim}},
\end{align}
where $\mu_{a,b}^\mathrm{sim}$ is the average of the first meeting times
obtained from the simulation.  We examine the average and the maximum of
the relative errors $\epsilon_{a,b}$ when changing starting node $b$
while the starting node $a$ is fixed.

Figure~\ref{fig:avg_degree-error} shows the average and the maximum of
the relative errors $\epsilon_{a,b}$ of the expected first meeting
time $\mu_{a,b}$ in the BA and ER graphs with different settings of
the average weighted degree $d_\mathrm{avg}$.  In this figure, we do
not plot the results for the ER graphs with $d_\mathrm{avg} = 2$ and
$4$, because a connected ER graph cannot be generated.  According to
Fig.~\ref{fig:avg_degree-error}, if $d_\mathrm{avg} \ge 4$, then the
maximum of relative errors $\epsilon_{a,b}$ is only a few
percent.  Therefore, the spectral
formula~\eqref{eq:avg_first_meeting_time_pre} is valid for the graphs
with $d_\mathrm{avg} \ge 4$.

\insertfig{avg_degree-error}{Average weighted degree $d_\mathrm{avg}$ vs. the average and the maximum of the relative error $\epsilon_{a,b}$ for the expected first meeting $\mu_{a,b}$}

We discuss the reason that the relative error $\epsilon_{a,b}$ is
large when we use a BA graph with a small-average weighted
degree~(i.e., $d_\mathrm{avg} = 2$).  In Sect.~\ref{sec:analysis}, we
use Eq.~\eqref{eq:meeting_prob} to derive the spectral
formula~\eqref{eq:avg_first_meeting_time_pre}.
Equation~\eqref{eq:meeting_prob} assumes that the first meeting
probability of two random walkers at node $c$ is proportional to
$d_{c}^2$.  Hence, we confirm the acceptance of this assumption to
clarify the reason for the large relative error.

Figures~\ref{fig:deg-prob_meettime_ba}~(a) through (c) show scatter
plots of the first meeting frequency of two random walkers at node $c$
in BA graphs with different settings of the average weighted degree
$d_\mathrm{avg}$.  The first meeting frequency at each node was
obtained from the simulation, where the starting nodes $a$ and $b$ are
fixed.  In order to confirm easily the correctness of the assumption,
we plot the fitting curve of $d_c^2$ in these figures.  According to
Figs.~\ref{fig:deg-prob_meettime_ba}~(a) through (c), the first
meeting frequencies with $d_\mathrm{avg} = 2$ differ only largely from
the fitting curve, with the consequence that the assumption must not
be accepted for the cases with $d_\mathrm{avg} = 2$.  Therefore, we
conclude that the large relative error shown in
Fig.~\ref{fig:avg_degree-error} is caused by the assumption for
Eq.~\eqref{eq:meeting_prob}.

\insertfigthree[.3]{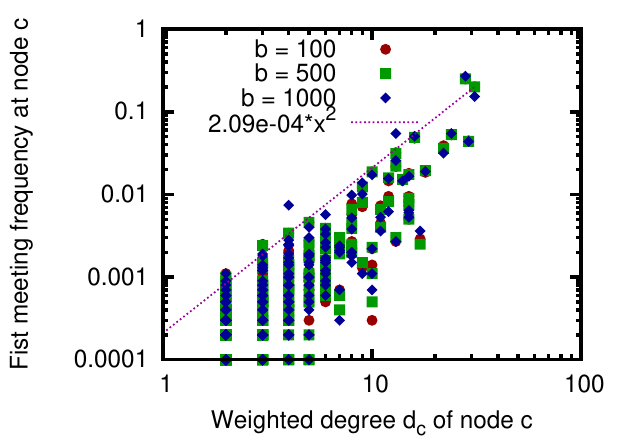}{$d_\mathrm{avg} = 2$}
{deg-prob_firstmeet_ba_N=1000_m0=4_m=2_src1=1}{$d_\mathrm{avg} = 4$}
{deg-prob_firstmeet_ba_N=1000_m0=6_m=3_src1=1}{$d_\mathrm{avg} = 6$}
{deg-prob_meettime_ba}{Weighted degree $d_c$ of node $c$ vs. the first meeting frequency at node $c$ in BA graphs}

According to the results, the spectral formula of first meeting time
$\mu_{a,b}$ will be valid if the average weighted degree
$d_\mathrm{avg}$ is sufficiently large~(i.e., $d_\mathrm{avg} \ge 4$).

\subsection{Validity for the Principal Component of the Expected First Meeting Time $\mu_{a,b}$}

We clarify the validity for the principal component of the expected
first meeting time $\mu_{a,b}$ derived in
Sect.~\ref{sec:analysis}.  Specifically, we examine the relative error
$\epsilon_{a,b}'$ of the approximation formula~\eqref{eq:approx_meet}
obtained when the expected first meeting time $\mu_{a,b}$ is given by
the principal component~(i.e., $s_1^2/s_2$).  The relative error
$\epsilon_{a,b}'$ is defined as
\begin{align}
 \epsilon_{a,b}' := \frac{\left|\frac{s_1^2}{s_2} - \mu_{a,b}^\mathrm{sim}\right|}{\mu_{a,b}^\mathrm{sim}}.
\end{align}

Figures~\ref{fig:N-err_in_meet_ba} and \ref{fig:N-err_in_meet_er} show
the averages of the relative errors $\epsilon_{a,b}'$ of the
approximation formula~\eqref{eq:approx_meet} for BA and ER graphs with
different numbers of nodes, $n$, respectively. The average of the
relative errors $\epsilon_{a,b}'$ was calculated from 10,000
simulations, where the starting nodes $a$ and $b$ are selected
randomly. In Fig.~\ref{fig:N-err_in_meet_er}, we do not plot the
result for $n = 10,000$ and $d_\mathrm{avg} = 6$, because a connected
ER graph cannot be generated.  According to the results, if the
average weighted degree $d_\mathrm{avg}$ is sufficiently large, the
relative error $\epsilon_{a,b}'$ is small, and the derived principal
component is valid.  This can also be explained by
Eq.~\eqref{eq:err_in_meet}.  The right-hand side of
Eq.~\eqref{eq:err_in_meet} represents the upper bound of the error in
the approximation formula~\eqref{eq:approx_meet}.  If the average
weighted degree $d_\mathrm{avg}$ is large, the minimum weighted degree
$d_{\rm min}$ is also large.  As the minimum weighted degree $d_{\rm
  min}$ increases, the upper bound becomes small, and the relative
error $\epsilon_{a,b}'$ of the approximation
formula~\eqref{eq:approx_meet} can be expected to decrease.  Moreover,
according to Figs.~\ref{fig:N-err_in_meet_ba} and
\ref{fig:N-err_in_meet_er}, the average of the relative errors
$\epsilon_{a,b}'$ is constant or becomes smaller as $n$ increases, and
hence the approximation formula~\eqref{eq:approx_meet} is also
effective for large-scale graphs.

\insertfig{N-err_in_meet_ba}{Number of nodes, $n$, vs. the average of the relative errors $\epsilon_{a,b}'$ of the approximation formula~\eqref{eq:approx_meet} in the BA graphs}
\insertfig{N-err_in_meet_er}{Number of nodes, $n$, vs. the average of the relative errors $\epsilon_{a,b}'$ of the approximation formula~\eqref{eq:approx_meet} in the ER graphs}

From the above results, the derived principal component is valid if
the average weighted degree $d_\mathrm{avg}$ is sufficiently
large~(i.e., $d_\mathrm{avg} \ge 4$).  According to the
site~\cite{KONECT}, which collecting statistical information~(e.g.,
average degree) of various existing graphs, the average degree of a
typical graph is greater than four.  Hence, our analysis results are
expected to be useful for many real graphs.

Finally, we confirm the effect of the graph structure on the expected
first meeting time $\mu_{a,b}$ revealed in
Sect.~\ref{sec:analysis}.  According to the approximation
formula~\eqref{eq:approx_meet}, $\mu_{a,b}$ increases as $s_1^2/s_2$
increases.  To confirm the effect from the numerical example, we
compare $s_1^2/s_2$ and the average of the first meeting times
obtained in the simulation.

Figures~\ref{fig:sd-meet_ba} and \ref{fig:sd-meet_er} show the
averages of the first meeting times obtained from the simulation with
different settings of $s_1^2/s_2$ in the BA and ER graphs,
respectively.  To calculate the average of the first meeting times, we
conduct 10,000 simulations, where the starting nodes $a$ and $b$ are
selected randomly.  In these figures, we plot the straight line for $y
= x$ to confirm the effect easily.  According to the results, the
average of the first meeting times is approximately along the $y = x$
line, except for the result for BA graphs with average weighted degree
$d_\mathrm{avg} = 2$.  Therefore, the effect is also confirmed from
the numerical example if the average weighted degree $d_\mathrm{avg}$
is sufficiently large.

\insertfig{sd-meet_ba}{Principal component $s_1^2/s_2$ vs. the average of the first meeting times in the BA graphs}
\insertfig{sd-meet_er}{Principal component $s_1^2/s_2$ vs. the average of the first meeting times in the ER graphs}

\section{Conclusion and Future Work}
\label{sec:conclusion}

In this paper, we analyzed the first meeting time of multiple random
walks in arbitrary graphs and clarified the effects of graph
structures on its expected value.  First, we derived the spectral
formula of the expected first meeting time for two random walkers
using spectral graph theory.  Then, we examined the principal component
of the expected first meeting time using the derived spectral
formula.  The clarified principal component reveals that (a)~the
expected first meeting time is almost dominated by $n/(1+d_{\rm
  std}^2/d_{\rm avg}^2)$, and (b)~the expected first meeting time is
independent of the starting nodes of random walkers, where $n$ is the
number of nodes.  $d_{\rm avg}$ and $d_{\rm std}$ are the average and the
standard deviation of the weighted degree, respectively.  The
characteristic~(a) is useful for understanding the effect of the graph
structure on the first meeting time.  In addition, we confirmed the
validity of the analysis results through numerical examples.  According
to the revealed effects of the graph structures, the variance of the
coefficient for weighted degrees, $d_{\rm std}/d_{\rm avg}$~(degree
heterogeneity), facilitates the meeting of random walkers.

As future work, we plan to examine the validity of the analysis
results with real graphs and apply them to the development of
efficient graph algorithms.

\section*{Acknowledgment}

This work was supported by JSPS KAKENHI Grant Number 19K11927.

\bibliographystyle{IEEEtran}
\bibliography{bib/random_walk,bib/bib}

\profile{Yusuke Sakumoto} {received M.E. and Ph.D. degrees in the
  Information and Computer Sciences from Osaka University in 2008 and
  2010, respectively.  From 2010 to 2019, he was a associate professor
  of Tokyo Metropolitan University.  He is currently an associate
  professor at Kwansei Gakuin University.  His research work is in the
  area of analysis of communication network and social network.  He is
  a member of the IEEE, IEICE and IPSJ.}

\profile{Hiroyuki Ohsaki}{received the M.E. degree in the Information
  and Computer Sciences from Osaka University, Osaka, Japan, in 1995.
  He also received the Ph.D. degree from Osaka University, Osaka,
  Japan, in 1997. He is currently a professor at Department of
  Informatics, School of Science and Technology, Kwansei Gakuin
  University, Japan. His research work is in the area of design,
  modeling, and control of large-scale communication networks. He is a
  member of IEEE and Institute of Electronics, Information, and
  Computer Engineers of Japan (IEICE).}

\end{document}